\documentclass[letterpaper,twocolumn,aps,prb,superscriptaddress,floatfix]{revtex4}

\usepackage{graphicx}

\begin{document}

\title{Influence of
intermartensitic transitions on transport properties of
$\mathbf{Ni}_{2.16}\mathbf{Mn}_{0.84}\mathbf{Ga}$ alloy}

\author{V.~V.~Khovailo}
\author{K.~Oikawa}
\affiliation{National Institute of Advanced
Industrial Science and Technology, Tohoku Center, Sendai
983--8551, Japan}

\author{C.~Wedel}
\affiliation{Institute of Multidisciplinary Research for Advanced
Materials, Tohoku University, Sendai 980--8577, Japan}

\author{T.~Takagi}
\affiliation{Institute of Fluid Science, Tohoku University, Sendai
980--8577, Japan}

\author{T.~Abe}
\affiliation{National Institute of Advanced Industrial Science and
Technology, Tohoku Center, Sendai 983--8551, Japan}

\author{K.~Sugiyama}
\affiliation{Earth and Planetary Science, Graduate School of
Science, The University of Tokyo, Tokyo 113--0033, Japan}

\begin{abstract}
Magnetic, transport, and x-ray diffraction measurements of
ferromagnetic shape memory alloy Ni$_{2.16}$Mn$_{0.84}$Ga revealed
that this alloy undergoes an intermartensitic transition upon
cooling, whereas no such a transition is observed upon subsequent
heating. The difference in the modulation of the martensite
forming upon cooling from the high-temperature austenitic state
[5-layered (5M) martensite], and the martensite forming upon the
intermartensitic transition [7-layered (7M) martensite] strongly
affects the magnetic and transport properties of the alloy and
results in a large thermal hysteresis of the resistivity $\rho$
and magnetization $M$. The intermartensitic transition has an
especially marked influence on the transport properties, as is
evident from a large difference in the resistivity of the 5M and
7M martensite, $(\rho _{\mathrm{5M}} - \rho _{\mathrm{7M}})/\rho
_{\mathrm{5M}} \approx 15\%$, which is larger than the jump of
resistivity at the martensitic transition from the cubic
austenitic phase to the monoclinic 5M martensitic phase. We assume
that this significant difference in $\rho$ between the martensitic
phases is accounted for by nesting features of the Fermi surface.
It is also suggested that the nesting hypothesis can explain the
uncommon behavior of the resistivity at the martensitic
transition, observed in stoichiometric and near-stoichiometric
Ni-Mn-Ga alloys.
\end{abstract}


\maketitle

\section{Introduction}

Intermetallic compounds undergoing thermoelastic martensitic
transformation when in the ferromagnetic state (ferromagnetic
shape memory alloys) have attracted considerable interest (see,
for a recent review, Ref.~\onlinecite{0-V}). This is due to the
fact that they exhibit large magnetic-field-induced strains which
can be obtained either by re-orientation of martensitic
variants~\cite{1-U,2-M} or by shifting the martensitic transition
temperature~\cite{3-C,4-T}. In addition to this effect of
practical significance, the ferromagnetic shape memory alloys have
been the subject of numerous studies due to their rich phase
diagrams. In particular, some of these alloys exhibit several
phase transitions between different crystallographic modifications
of martensite, induced by a change of composition, temperature or
stress, or by the combination of these parameters.

A prototype of the ferromagnetic shape memory alloys, Ni$_2$MnGa,
is a representative of Mn-containing Heusler alloys. It orders
ferromagnetically at Curie temperature $T_C = 376$~K. Upon cooling
down to $T_m = 202$~K it undergoes a reversible thermoelastic
martensitic transformation from the Heusler (L$2_1$) cubic
structure to a roughly tetragonal crystal structure. Both $T_m$
and $T_C$ are sensitive to stoichiometry. For instance, a partial
substitution of Mn for Ni in Ni$_{2+x}$Mn$_{1-x}$Ga alloys results
in an increase of $T_m$ and a decrease of $T_C$ until they couple
in a composition range $x = 0.18 - 0.20$ (Ref.~\onlinecite{5-V}).

An early neutron diffraction study~\cite{6-W} of the martensitic
structure of stoichiometric Ni$_2$MnGa showed that along with
strong tetragonal reflections there were several additional peaks
on the diffraction pattern. Based on this observation, the authors
suggested that the martensitic phase has a modulated crystal
structure. Further studies revealed~\cite{note} that modulation
and, therefore, the crystal structure of the martensite forming
from the parent austenitic phase, depends on composition
(Ref.~\onlinecite{7-P} and references therein). By now, five- and
seven-layered martensitic phases modulated along the
$(110)[1\overline{1}0]$ system and a non-modulated martensitic
phase have been established to exist in Ni-Mn-Ga alloys. In
addition, the observation of longer-period modulations of the
martensite has been reported~\cite{8-C}.

The crystal structure of martensite was found to be very unstable
to the application of external stresses~\cite{9-K,10-M,11-V,12-M}.
It turned out that the sequence of stress-induced
martensite-martensite transformations depends on many factors,
such as the composition of the sample, temperature of the test,
and the crystallographic direction along which the stress was
applied. Besides composition- or stress-induced changes in the
crystal structure of martensite, some off-stoichiometric Ni-Mn-Ga
alloys undergo a sequence of temperature-induced
martensite-martensite phase transitions. Apart from Ni-Ti
(Ref.~\onlinecite{13-L} and references therein) and
Ni$_{50}$Mn$_{50-x}$Al$_x$ (Ref.~\onlinecite{13a-I}) systems,
temperature-induced intermartensitic transitions have not been
observed in other shape memory alloys.

In Ni-Mn-Ga intermartensitic transitions are, as evident from
calorimetric measurements~\cite{14-C}, first-order phase
transitions. As compared with the martensitic transformation, the
intermartensitic transitions exhibit several distinctive features.
They are a large, exceeding 100 K, temperature hysteresis and a
considerable difference in transport properties between the
martensitic phases involved in an intermartensitic
transition~\cite{15-C,16-W,17-S,18-S,19-W,20-K,20q-L,20r-C}.
Transport measurements of Ni-Mn-Ga alloys undergoing
intermartensitic transitions~\cite{5-V,19-W,20-K,20q-L} have
indicated that the difference in the resistivity between
martensitic phases is comparable or even larger than that observed
at the martensitic transformation temperature. This seems to be
unusual because martensitic transformation has a stronger
influence on the physical characteristics (crystal structure,
Fermi surface, magnetic properties) of the materials.

Since these features of intermartensitic transitions have not been
discussed earlier we have studied and analyzed the transport
properties of Ni$_{2.16}$Mn$_{0.84}$Ga undergoing an
intermartensitic transition. In our study we have also used a
stoichiometric Ni$_2$MnGa sample, prepared by the same method as
Ni$_{2.16}$Mn$_{0.84}$Ga.

\section{Experimental}

A polycrystalline ingot of Ni$_{2.16}$Mn$_{0.84}$Ga composition
was prepared by arc melting high purity constituent elements in
argon atmosphere. In order to get a good compositional
homogeneity, the ingot was re-melted several times and annealed at
1050~K for nine days with subsequent quenching in ice water.
Samples for resistivity and magnetization measurements were cut
from the middle part of the ingot. Temperature dependencies of
resistivity and magnetization were measured with a heating/cooling
rate of 1~K/min by a standard four-probe technique and by a
vibrating sample magnetometer, respectively. The crystal structure
of the alloy was examined using a Philips X-Pert system in a wide
temperature interval. For the powder x-ray diffraction
measurements, part of the ingot was crushed into a fine powder.
The powder was sealed in an evacuated quartz tube and annealed at
1050~K for five days in order to remove residual stress and
improve the peak shape of diffraction patterns.

\section{Experimental results}

The temperature dependencies of electrical resistivity of
Ni$_{2.16}$Mn$_{0.84}$Ga, measured upon cooling and heating, are
shown in Fig.~1. Cooling from high temperatures results in the
formation of a long-range ferromagnetic ordering at $T_C = 337$~K
which is accompanied by a change in the slope of the resistivity
curve due to the decrease in electron-magnon scattering. The
jump-like increase of the resistivity at $T_m \approx 309$~K
corresponds to the transition from the high-temperature austenitic
to a low-temperature martensitic phase.

Besides the change in the slope of the curve at $T_C = 337$~K and
the jump-like increase of $\rho$ at $T_m = 309$~K, one more marked
change in the slope of the cooling curve is observed at $T_I =
283$~K. Since this anomaly is observed when the sample is in the
martensitic state, this means that a martensite-martensite
transformation occurs in Ni$_{2.16}$Mn$_{0.84}$Ga. Based on the
results of transmission electron microscopy (TEM) observation of a
sample of this composition~\cite{21-W}, which revealed that the
crystal structure of martensite in Ni$_{2.16}$Mn$_{0.84}$Ga is
characterized by a five-layered modulation (5M) at room
temperature and seven-layered modulation (7M) at $T = 173$~K, we
conclude that the anomaly of $\rho$ at $T_I = 283$~K corresponds
to the onset of intermartensitic transition from a five- to
seven-layered martensite (5M $\to$ 7M). Cooling the sample below
$T_I$ initially results in a distinct decrease of the resistivity,
which becomes less temperature-dependent on further cooling. No
anomaly corresponding to the end of the 5M $\to$ 7M
intermartensitic transition was observed on the resistivity curve
in the temperature interval of the measurements. The absence of
such an anomaly implies that the fraction of the 5-layered
martensite gradually decreases with decreasing temperature and
therefore both the 5M and 7M martensitic phases coexist over a
wide temperature interval. Subsequent heating revealed a
monotonous increase of the resistivity up to the reverse
martensitic transformation temperature.

\begin{figure}[t]
\begin{center}
\includegraphics[width=8cm]{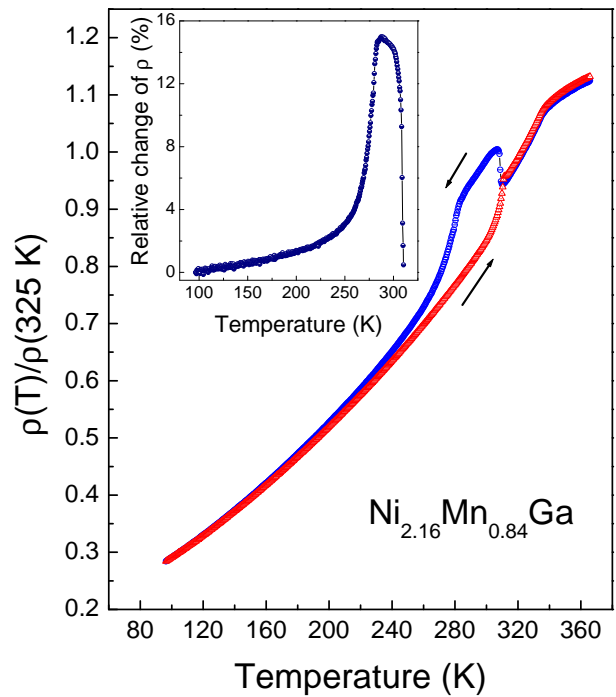}
\caption{Temperature dependencies of electrical resistivity for
Ni$_{2.16}$Mn$_{0.84}$Ga measured during cooling and heating. The
inset shows the difference between $\rho$ measured upon cooling
and heating, $(\rho _{cooling} - \rho _{heating})/\rho
_{cooling}$.}
\end{center}
\end{figure}

Since the 5M $\to$ 7M intermartensitic transition is not completed
in the studied temperature interval and because of the absence of
the reverse 7M $\to$ 5M intermartensitic transition upon
subsequent heating, the resistivity exhibits very large thermal
hysteresis. At temperatures below the martensitic transformation,
the heating curve deviates from the curve measured upon cooling,
and the difference between $\rho$ measured upon cooling and
heating progressively increases as the temperature is increased
(inset in Fig.~1). Assuming for simplicity that at $T = 100$~K
there exists only a tiny fraction of the 5M martensite, we can
estimate the difference in the resistivity between the 5M and 7M
phases, $\Delta \rho = (\rho _{\mathrm{5M}} - \rho
_{\mathrm{7M}})/\rho _{\mathrm{5M}}$. As is seen from the inset in
Fig.~1, $\Delta \rho \approx 15\%$ in a temperature interval from
283 to 300~K .

Due to the absence of the reverse intermartensitic transition upon
heating, the behavior of $\rho$ at the martensitic transformation
measured upon cooling and heating is quite different. Whereas
$\rho$ shows a jump-like increase during direct martensitic
transformation from the parent phase to the 5M martensite, the
resistivity steepens up when approaching reverse martensitic
transformation from the 7M martensite to the parent phase
(Fig.~1). If the anomaly of $\rho$ at $T_I = 283$~K indeed
corresponds to the onset of the intermartensitic transition, below
which the fraction of the 5M martensite gradually decreases, the
behavior of $\rho$ at martensitic transformation temperature $T_m$
should depend on the proportion of the 5M and 7M phases. In order
to check this, we measured several partial cooling-heating cycles.

\begin{figure}[t]
\begin{center}
\includegraphics[width=7cm]{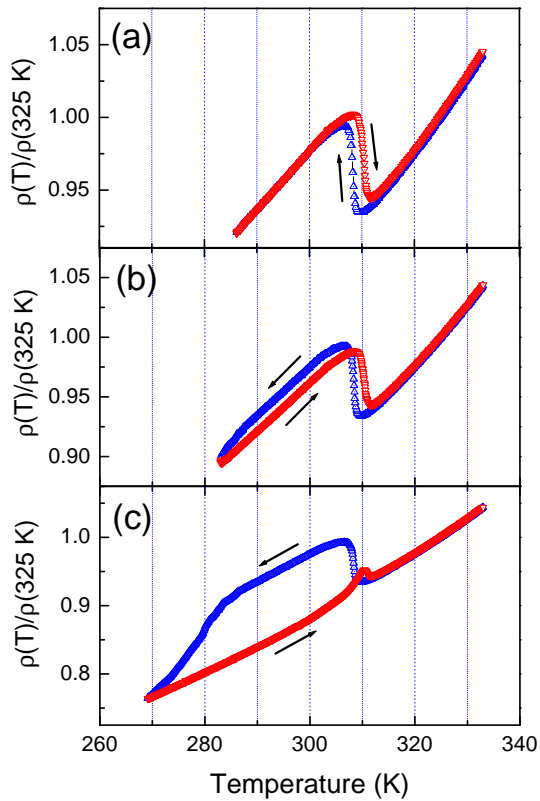}
\caption{Electrical resistivity of Ni$_{2.16}$Mn$_{0.84}$Ga
measured upon partial cooling-heating cycles in temperature
intervals 286 - 333 K (a), 282 - 333 K (b), and 269 - 333 K (c).}
\end{center}
\end{figure}

The results of these measurements (Fig.~2) indicate that the
behavior of $\rho$ at $T_m$ upon cooling is always the same
(jump-like increase), but that measured upon heating substantially
depends on the temperature, down to which the sample was cooled.
If the sample is cooled down to $T > T_I$, the resistivity upon
subsequent warming follows the cooling path and $\rho$ exhibits a
marked jump-like decrease during transformation to the austenitic
phase [Fig.~2(a)]. This means that cooling of the sample to $T =
286$~K, which is slightly higher than the $T_I$ temperature, did
not result in the formation of a two-phase state and the sample
remains in the 5M martensitic state upon subsequent heating.

When the sample is cooled somewhat below $T_I$, the behavior of
resistivity at $T_m$ upon heating is still similar to that
observed upon cooling. This is evident from Fig.~2(b), where the
sample was cooled to $T = 282$~K, i.e. 1~K below the $T_I$
temperature. This means that in the two-phase state of the sample,
attained by cooling slightly below $T_I$, the behavior of $\rho$
upon heating is governed by a considerably larger fraction of the
5M martensite. Note that in this case the heating curve is
parallel to the cooling curve, indicating that the two-phase state
existing at $T = 282$~K is preserved up to the reverse martensitic
transformation. In other words, the fraction of the 7M martensite
does not transform to the 5M martensite upon heating from 282~K.
This observation implies that the onset of the reverse
intermartensitic transition is above the martensitic
transformation temperature $T_m$.

Finally, when the sample is cooled down to $T = 269$~K, the
resistivity upon subsequent heating exhibits behavior, typical of
the 7M martensitic phase [Fig.~2(c)], and $\rho$ shows a small
kink at the martensitic transformation temperature. In a
temperature interval from 283~K to 309~K, the difference in $\rho$
between heating and cooling curves is $\sim 12$\%, indicating that
approximately 80\% of the 7M martensite had been formed upon
cooling to 269~K. Based on the results presented in Fig.~2 one can
conclude that the 7M martensite appears upon cooling below $T_I =
283$~K and the fraction of this martensitic phase considerably
exceeds that of the 5M martensitic phase at $T < 270$~K.

\begin{figure}[t]
\begin{center}
\includegraphics[width=7cm]{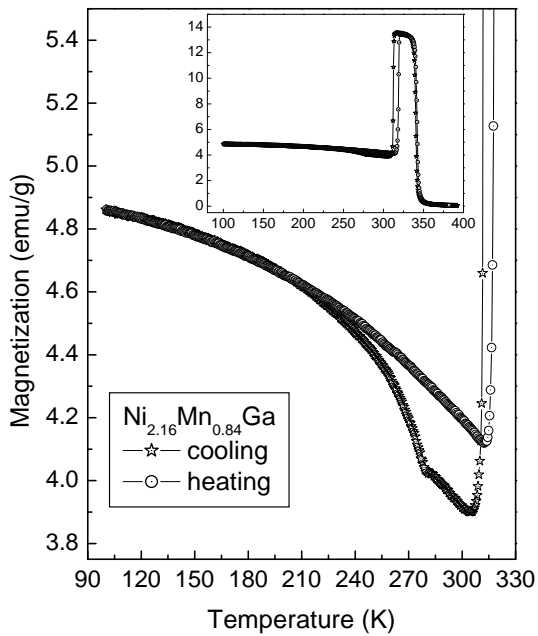}
\caption{Temperature dependencies of the magnetization of
Ni$_{2.16}$Mn$_{0.84}$Ga measured in a 0.1~T magnetic field. The
inset shows $M(T)$ over the entire temperature interval.}
\end{center}
\end{figure}

\begin{figure}
\begin{center}
\includegraphics[width=8cm]{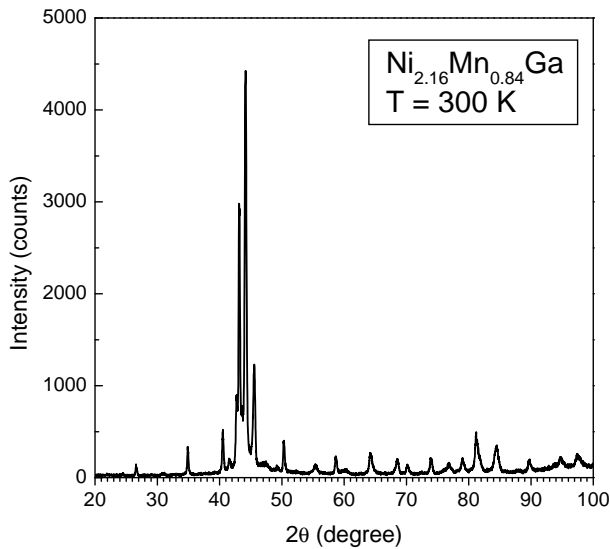}
\caption{X-ray diffraction pattern of the 5M martensitic phase of
Ni$_{2.16}$Mn$_{0.84}$Ga forming upon cooling from the
high-temperature austenitic phase.}
\end{center}
\end{figure}

The magnetization $M$ of Ni$_{2.16}$Mn$_{0.84}$Ga measured in a
0.1~T magnetic field is shown in Fig.~3. The Curie temperature,
determined from this measurement, is equal to 340~K (inset in
Fig.~3). The anomaly at $\approx 310$~K, exhibiting a temperature
hysteresis of $\sim 6$~K, corresponds to the martensitic
transformation. Like the resistivity, the magnetization of the
sample shows a large thermal hysteresis in the martensitic state.
A well-defined change in the slope of the $M(T)$ curve measured
upon cooling at $T = 279$~K corresponds to the onset of the
intermartensitic transition to the 7M phase. This characteristic
temperature, determined from the magnetization measurements, is
slightly lower than that obtained from the resistivity data. This
difference can be accounted for by the fact that $M(T)$ and $\rho
(T)$ measurements were performed on different samples. As is seen
from Fig.~3, in the 0.1~T magnetic field magnetization of the 5M
martensitic phase is lower than that of the 7M phase and the
difference between them gradually diminishes as the temperature is
lowered.

The thermal hysteresis of $M$ is observed only in low magnetic
fields. Measurements of $M(T)$ in a magnetic field of 1~T showed
no thermal hysteresis of $M$ in the martensitic state, which means
that both the martensitic phases have the same magnetization in
this field. Therefore, it can be concluded that magnetization
saturation of these two martensitic phases is the same.

The diffraction pattern of Ni$_{2.16}$Mn$_{0.84}$Ga, taken at room
temperature is shown in Fig.~4. To be sure that the measured
diffraction pattern corresponds to the 5M martensite, the powder
was heated above the martensitic transformation temperature $T_m$
and the measurement was performed on the powder cooled to room
temperature from the austenitic state. Preliminary analysis of the
room temperature diffraction pattern of Ni$_{2.16}$Mn$_{0.84}$Ga
showed that the crystal structure of the martensite formed upon
cooling from the austenitic phase can be interpreted as a
monoclinic one with lattice parameters $a = 0.42$~nm, $b =
0.55$~nm, $c = 2.10$~nm, and $\beta = 92^{\circ}$.

X-ray diffraction measurement at a lower temperature, $T = 77$~K,
confirmed the occurrence of the intermartensitic transition, seen
on the $\rho (T)$ and $M(T)$ curves. The crystal structure of the
7M martensitic phase was interpreted as monoclinic with lattice
parameters $a = 0.426$~nm, $b = 0.543$~nm, $c = 2.954$~nm, and
$\beta = 94.3^{\circ}$. Further cooling down to $T = 10$~K did not
result in a change of the diffraction pattern observed at $T =
77$~K. The results of these measurements are shown on an enlarged
scale in Fig.~5.

\section{Discussion}

The results of our resistivity measurements (Fig.~1) indicate that
different martensitic phases are considerably distinguished by
their transport properties, namely $\rho _{\mathrm{5M}}$ is larger
than $\rho _{\mathrm{7M}}$ by 15\%. Generally, this difference can
be caused by two factors: by changes in the scattering probability
and/or by changes in the electronic structure. Since both the
phases exhibit similar plate-like morphology~\cite{21-W} we
suggest that the 15\% difference in the resistivity of these
phases can not be accounted for by an increase in the scattering
centers. Therefore, the origin of this difference has to be looked
for in the Fermi surface features. Indeed, it is generally
acknowledged that the formation of a long-range ordering observed
in a large number of compounds is associated with the nesting
properties of the Fermi surface. This is true as for the case of
long-range structural ordering~\cite{22-Z,23-W}, as for the case
of long-range magnetic ordering, such as spin- or charge-density
waves~\cite{24-F,25-F,26-A,27-A}. The periodicity of long-range
ordering is determined by the nesting vector on the Fermi surface.

It is conceivable that the various martensitic phases forming in
Ni-Mn-Ga alloys are driven by the geometry of the Fermi surface
that has a nesting vector corresponding to the modulation of
martensite, as was suggested in~\cite{28-V}. This suggestion
implies that martensitic phases with different nesting vectors
have different fractions of nested Fermi surface. On the other
hand, it is well known that the nesting considerably affects the
transport properties of a metal due to the condensation of
electrons in the nesting parts of the Fermi surface. Therefore,
change in the modulation can affect the number of conduction
electrons $n_{\mathrm{eff}}$ due to the change of the Fermi
surface available for conduction.

In the simple relaxation time approximation

$$\rho = m^*/n_{\mathrm{eff}}\,e^2\tau ,$$

\noindent where $m^*$ is the effective mass, $e$ is the electronic
charge, and $\tau$ is the relaxation time. Assuming that the
relaxation time $\tau$ is the same for both 5M and 7M martensitic
phases, the experimental observation that $\rho _{5M}
> \rho _{7M}$ indicates that the 5M phase has a fewer number of
conduction electrons $n_{\mathrm{eff}}$ than the 7M phase. Note
that if the proposed explanation is valid, one can expect to
observe an anisotropic behavior of $\rho$ in a Ni-Mn-Ga single
crystalline sample, as is the case in Cr~\cite{24-F} and
heavy-fermion compounds~\cite{29-M,30-M,31-M}.

\begin{figure}[t]
\begin{center}
\includegraphics[width=7cm]{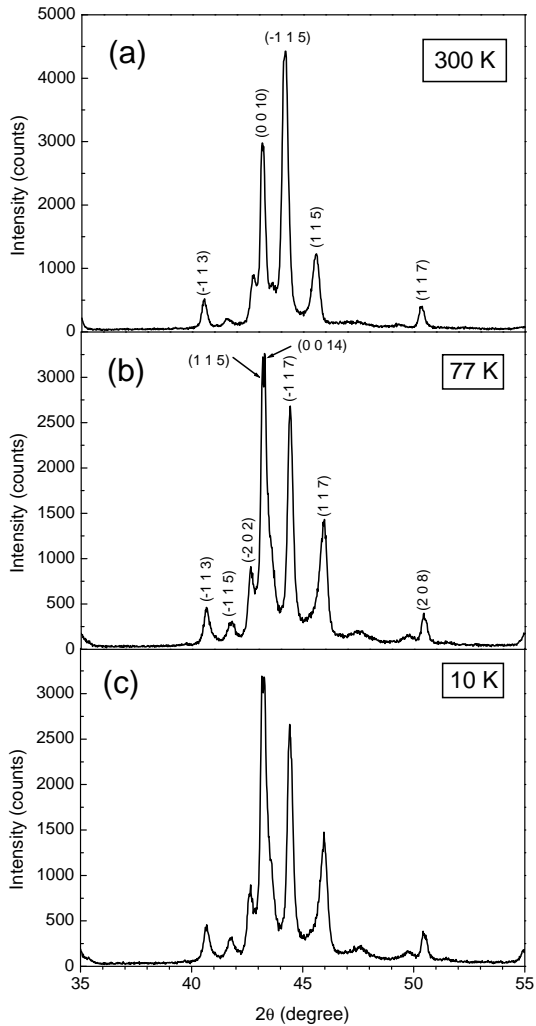}
\caption{Results of X-ray diffraction measurements of
Ni$_{2.16}$Mn$_{0.84}$Ga performed at different temperatures upon
cooling. The diffraction patterns were collected at 300~K (a),
77~K (b), and 10~K (c).}
\end{center}
\end{figure}

In the above discussion we did not consider the possibility that
electron-magnon scattering can be different in the 5M and 7M
martensites. However, according to Friedel and de
Genner~\cite{32-F}, temperature dependence of the magnetic
resistivity $\rho _{\mathrm{mag}}$ can be described as

$$\rho _{\mathrm{mag}} = \rho _{\mathrm{s-d}}[1-\sigma ^2(T)],$$

\noindent where $\rho _{\mathrm{s-d}}$ is the temperature
independent spin-disorder resistivity and $\sigma ^2(T) =
M_s(T)/M_s(0)$, $M_s(T)$ and $M_s(0)$ are magnetization
saturations at a finite temperature $T$ and at $T = 0$~K. Since
our magnetic measurements have shown that the magnetization
saturation $M_s$ of the 5M and 7M phases is the same, it can be
concluded that both the phases are characterized by the same
electron-magnon scattering.

It is apparent that, along with the unusual transport properties
of Ni$_{2.16}$Mn$_{0.84}$Ga discussed above, the proposed nesting
hypothesis can satisfactory explain the uncommon behavior of
$\rho$ at martensitic transformation temperature $T_m$, observed
in stoichiometric and near-stoichiometric Ni$_2$MnGa alloys.
Indeed, since martensitic transformation results in a drastic
change of crystal structure, Fermi surface, mean free path and so
on, one can expect to detect a well-defined anomaly at $T_m$,
which is indeed generally observed in shape memory
alloys~\cite{33}. In contrast to this, $\rho (T)$ measurements for
stoichiometric and near-stoichiometric Ni$_2$MnGa revealed only a
change in the slope of the curve at the martensitic transformation
temperature~\cite{5-V,34-K,35-C,36-Z}.

We argue that such a peculiar behavior of $\rho$ at $T_m$ in
Ni$_2$MnGa is due to the premartensitic transition, occurring
above $T_m$, despite the fact that the cubic symmetry of the
crystal structure is preserved upon this transition~\cite{37-Z}.
As is shown in Fig.~6, for the austenitic phase of stoichiometric
Ni$_2$MnGa $\rho (T)$ can be fitted by a $T^n$ dependence $(n
\approx 3)$. The experimental curve deviates from the fit at Curie
temperature $T_C = 376$~K and at $T = 266$~K which matches well
with the premartensitic transition temperature $T_P = 260$~K for
the stoichiometric composition~\cite{37-Z}. The driving force for
the premartensitic transition is believed to be Fermi surface
nesting, as was suggested by Zheludev \textit{et.
al.}~\cite{37-Z}, and recent theoretical calculation~\cite{38-L}
supports this hypothesis. The deviation of the resistivity from
the $T^n$ dependence below $T_P$ could be caused, for instance, by
an increase in the relaxation time $\tau$ due to the modulation of
the cubic structure in the premartensitic phase. In our opinion,
however, the primary role in this process is played by the
condensation of the conduction electrons in the nesting part of
Fermi surface. Assuming that without the premartensitic transition
$\rho (T)$ would follow the $T^n$ dependence down to the
martensitic transformation temperature $T_m = 202$~K, it can be
concluded from Fig.~6 that in this case the difference in $\rho$
between the high temperature austenitic and low temperature
martensitic phases is significant and the resistivity should
exhibit a jump-like behavior, as in other shape memory alloys or
in off-stoichiometric Ni-Mn-Ga.

\begin{figure}[t]
\begin{center}
\includegraphics[width=8cm]{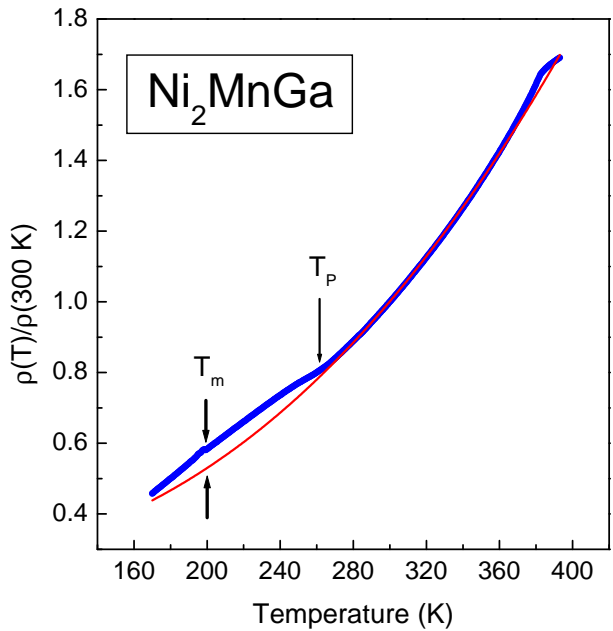}
\caption{Temperature dependence of electrical resistivity for the
stoichiometric Ni$_2$MnGa. The solid line is a fit to the
experimental curve.}
\end{center}
\end{figure}

With deviation from the stoichiometry, the martensitic transition
temperature increases or decreases, depending on substitution,
whereas the $T_P$ temperature is less composition
dependent~\cite{39-G,40-K,41-Z}. In the case of increasing $T_m$
this leads to the disappearance of the premartensitic transition
in a critical composition and, as a result, in off-stoichiometric
Ni-Mn-Ga alloys a marked jump-like behavior of $\rho$ is observed.

\section{Conclusion}

Temperature-induced intermartensitic transitions observed in
certain Ni-Mn-Ga alloys give rise to an anomalously large thermal
hysteresis of magnetic and transport properties, which is not
observed in other compounds. This thermal hysteresis is accounted
for by the coexistence of both martensitic phases in a wide
temperature interval. As is evident from the resistivity
measurements of Ni$_{2.16}$Mn$_{0.84}$Ga, the difference in $\rho$
between 5M and 7M martensite is about 15\%, which is even larger
than that observed upon the martensitic transformation. We have
suggested that such a significant difference is accounted for by
the geometry of the Fermi surface that has a different nesting
vector in 5M and 7M martensitic phases. If this assumption is
valid, an anisotropic behavior of $\rho$ in a Ni-Mn-Ga single
crystal of the same or similar composition can be reasonably
expected. Therefore, further studies of single crystalline samples
are required for better understanding structural instability of
various martensitic phases in Ni-Mn-Ga alloys.

In the framework of the nesting hypothesis we have also discussed
the peculiar behavior of $\rho$ at the martensitic transformation
temperature $T_m$ in stoichiometric Ni$_2$MnGa. We have argued
that this behavior of $\rho$ is caused by the condensation of
conduction electrons in the nesting part of the Fermi surface
occurring upon the premartensitic transition.

\bigskip


This work was partially supported by an Industrial Research Grant
Program in 2002 from New Energy and Industrial Technology
Development Organization (NEDO). We are thankful to Dr~M~Isobe for
the help with X-ray diffraction measurements. One of the authors
(V~V~K) gratefully acknowledges the Japan Society for the
Promotion of Science (JSPS) for a Fellowship Award.


\end{document}